\documentstyle[twoside,fleqn,espcrc2,epsfig]{article}
\newcommand\lr[1]{{\left({#1}\right)}}

\newcommand{\bt}{\begin{tabular}}
\newcommand{\et}{\end{tabular}}
\newcommand{\bd}{\begin{displaymath}}
\newcommand{\ed}{\end{displaymath}\noindent}
\newcommand{\ec}{\end{center}}
\newcommand{\bc}{\begin{center}}
\newcommand{\nn}{\nonumber\\}

\newcommand{\g}{\gamma}

\newcommand{\AmS}{{\protect\the\textfont2
  A\kern-.1667em\lower.5ex\hbox{M}\kern-.125emS}}

\def\beqa{\begin{eqnarray}}
\def\eqa{\end{array}}
\def\beqar{\begin{array}}
\def\eqa{\end{eqnarray}}
\def\bars{\begin{eqnarray*}}
\def\ears{\end{eqnarray*}}

\def\beqa{\begin{eqnarray}}
\def\eqa{\end{array}}
\def\beqar{\begin{array}}
\def\eqa{\end{eqnarray}}
\def\bars{\begin{eqnarray*}}
\def\ears{\end{eqnarray*}}

\title{On the QCD dipole content of hard photon and gluon probes}

\author{R.Peschanski\address{
{\it Service de Physique Th\'eorique}, CEA\ \ CEA-Saclay, F-91191 Gif-sur-Yvette Cedex,
France}}        
\begin{document}

\begin{abstract}
A gluon forward jet playing the r\^ole of a deep probe in high energy scattering, we analyze its infinite momentum QCD wave function in terms 
of dipole (color-siglet $q \bar q$) configurations using $k_T$-factorization properties. The comparison is made with virtual photon $q \bar q$ configurations. Some implications for hard  processes with forward jets at Hera and Tevatron are suggested.
\end{abstract}

\maketitle

\section{Photons and gluons as hard probes}

The studies on deep-inelastic scattering experiments by virtual photon probes have emphazised the interest of using the formalism of $q \bar q$ wave-functions of the photon in the infinite momentum frame \cite {bj71}. Indeed, the theoretical predictions for small-x structure functions at the leading-log approximation  can be expressed in terms of the probability of finding the $q \bar q$ configurations inside the photon probe \cite {bj71,ni94} convoluted with the high-energy dipole-dipole cross-section \cite {mu94,na95}. This is to be compared  to the $k$-factorization scheme \cite {ca91} where the photon vertex function is convoluted with  the unintegrated off-shell gluon structure function solution of the Balitskii, Fadin, Kuraev, Lipatov (BFKL) equation \cite {bfkl}. Interestingly enough, both schemes can be shown equivalent \cite {mu98}. More  recently, the photon wave-functions proved to be very useful in the investigations of the so-called ``hard diffraction'' processes characterized by a large rapidity gap between particles produced at the photon and proton vertices at Hera. Indeed, the physical process in which a color singlet $q \bar q$ configuration of the photon interacts \cite {ni94,bi96} with the proton seems to give an interesting and original insight on the dynamics of ``hard diffraction'' processes initiated by a photon probe. The interaction of various configurations can also be treated in the dipole model framework [7-9] using the  evolution with energy of the initial  color singlet $q \bar q$ configuration of the photon into a cascade of colorless dipoles.

In other reactions of interest, the hard probe is not furnished by a virtual photon. This is the case at the Tevatron for instance, where an energetic forward jet with high transverse momentum $k_T$ is used  as the hard QCD probe in various studies such as the Mueller-Navelet process \cite {mu86} or  ``hard diffraction'' processes at Tevatron. The gluon jet as a hard probe is also present in forward jet studies at Hera where it allows for  a completely perturbative QCD prediction for the hard gluon {\it vs.} hard photon scattering at high energy where one looks for a clear signal of the BFKL Pomeron
[11-14].  However, if a treatment of the gluon vertex following the $k$-factorization scheme is well known, the corresponding   formalism using
color-singlet $q \bar q$ wave-functions has not been either derived or even discussed. In fact, there exist some reasons why it is interesting to discuss
a hard probe using this formalism: first, it provides  a description of the hard probe in the {\it configuration space}. This configuration space is spanned by the variables $\vec r \equiv  r e^{i\phi}$, $z$ where $\vec r$ is the transverse      vector coordinate between $q$ and $\bar q$ and of the energy fraction $z$ (resp. $1\!-\!z$) brought by the quark (resp. the antiquark). Second, it is interesting to check the validity and limitations of the vertex factorization properties in configuration space. Third, ``hard diffraction'' has been formulated using the wave-function formalism and leads to quite non-trivial formulae (including 
the presence of interference terms [7,9]) which are not yet known for the gluon jet probe, e.g. at the Tevatron.

We present here a report on a new~\cite{pe99}  derivation and  discussion of the   color-singlet dipole content of a gluon jet probe. The starting point of this work starts from  the result of the color-singlet factorization scheme for the photon and its equivalence with the $k$-factorization scheme in deep inelastic scattering. Using a parallel to the photon approach for the color singlet content of a hard gluon jet, the configuration space distributions are  derived in terms of a definite combination of squared of Bessel functions depending on the variable $rk_T,$ where $k_T$ of the jet gives the hard scale. The validity of the color-singlet scheme is then checked both for the Mueller-Navelet \cite {mu86} and DIS forward jet formulae including the azimuthal assymetries \cite {del1,del2}. The resulting functions can be  discussed in terms of absorption of the gluon partial waves in the dipole center-of-mass frame. Some distinctive (w.r.t. a photon probe) features of the resulting states, such as the presence of both virtual and real contributions and the long distance behaviour of the wave-functions are typical features of the gluon jet which may have a significant impact on the phenomenology of hard reactions at high energy initiated by a forward jet probe.

\section{Color-singlet {\it versus} $k$-factorization at the photon vertex}

As mentionned in the introduction, the factorization properties of (resummed at the leadings logs) perturbative QCD in the high-energy regime can be put into two equivalent forms \cite {mu98}.  One form is using the $k$-factorization property \cite {ca91} which relates the $\gamma^*$-dipole cross-section to the product of the impact factors $V_{T,L}$ by a $g^*$-dipole cross-section where $g^*$ is  an off-mass-shell gluon ($T,L$ are for the $Transverse, Linear$ polarizations of the photon). The equivalent form is obtained through the photon wave-function in terms of $q\bar q$ configurations~\cite{bj71} convoluted with the $q\bar q$-dipole cross-section. The target dipole is considered to be small (i.e. massive) in order to justify the (resummed) perturbative QCD calculations. This equivalence is sketched in Fig.1, taking as an example the transverse helicity contributions from  the virtual photon.

\begin{figure}[htb]
\vspace{9pt}
\begin{center}
\epsfig{file=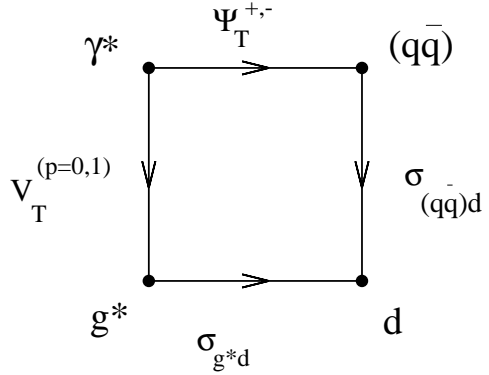,height=5cm}
\end{center}
\caption{{\it The two factorization schemes of the $\gamma ^*$-dipole cross-section.}
First Scheme: $k$-factorization  of the  gluon-dipole  impact factors $V_T^{(p=0,1)}$ where $p$ is for the conformal spin see Ref. \cite{ma99} for details and notations); Second Scheme: wave-function factorization of the $(q\bar q)\!-\!d$ cross-section. }
\label{fig:factorization}
\end{figure}

The light-cone wave functions of the photon for photon helicity ${\pm}1,0$ are~\cite{bj71} 
\begin{eqnarray}
\Psi^+_T \lr{z,r,Q^2}  &=& C\  z\ e^{i\varphi} \hat Q  K_1\lr{\hat Q r} 
\nonumber\\
\Psi^-_T \lr{z,r,Q^2}  &=& C\   \lr{1\!-\!z}\ e^{-i\varphi} \hat Q  K_1\lr{\hat Q r}
 \nonumber\\
\Psi_L \lr{z,r,Q^2}  &=& C\  2 \sqrt{z(1\!-\!z)}\ \hat Q  K_0\lr{\hat Q r} ,
\label{photon}
\end{eqnarray}
where $K_{0,1}$ are the known Bessel functions of second kind. By definition $\hat Q \equiv Q\sqrt{z(1\!-\!z)}$ and  the normalization is
$C^2\equiv \frac{\alpha_{em}N_c e_q^2}{4\pi\alpha_s}$ with $e_q$ the quark charge.

For a given target $t$ (e.g. the massive dipole $d$ in Fig.1), the transverse and longitudinal photon-target total cross sections $\sigma^t_{T,L}$ obtained by high-energy QCD in the leading logs approximation
 read
\beqa
\sigma^{\g}_{T,L}=\int d^2r dz\;\int d^2r_t dz_t\ \times
\nonumber\\
\; 
\Phi_{T,L}(r,z;Q^2)\Phi^t(r_t,z_t;Q_t^2)\>\sigma_{d}(r,r_t;Y)\ ,
\label{eqn:ftl}
\eqa
where, by definition,  $\Phi_T\equiv \vert\Psi^+_T\vert^2+\vert\Psi^-_T\vert^2 $ for the transversely polarized photon, $\Phi_L\equiv \vert\Psi_L\vert^2$ for the longitudinally polarized one and  $\Phi^t(r_t,z_t)$ defines the probability distribution of initial
 dipoles inside the target (e.g. $\delta(r_t\!-\!1/Q_t) \delta(z_t\!-\!1/2)$ for a massive dipole at scale $Q_t$).

Concerning the $q\bar q$ dipole cross section
$\sigma_d,$ it can be expressed using a Mellin transform and
reads~\cite{mu98}:
\beqa
\sigma_{d}(r,r_t;Y)=4\pi \int\frac{d\gamma}{2i\pi}
\left(r^2\right)^{1-\gamma}\left(r_t^2\right)^{\gamma}\nonumber\\
\times \ e^{\frac{\alpha_s N_c}{\pi}\chi(\gamma)Y}A_{el}(\gamma)\ ,
\label{dipoles}
\eqa
where  
\begin{equation}
\chi(\gamma)=2\Psi(1)-\Psi(\gamma)-\Psi(1\!\!-\!\!\gamma)\ ,
\label{eqn:kernel}
\end{equation}
stands for the BFKL kernel \cite{bfkl}
\begin{equation}
A_{el}(\gamma)=\frac{\alpha_s^2}{16\gamma^2(1\!-\!\gamma)^2}
              =\alpha_s^2 v(\gamma)v(1\!-\!\gamma)\ 
\label{eqn:27}
\end{equation}
is the first order elastic two-gluon exchange amplitude 
while the factor
\begin{equation}
v(\gamma)=\frac{2^{-2\gamma-1}}{\gamma}\frac{\Gamma(1\!-\!\gamma)}{\Gamma(1\!+\!\gamma)}
\label{eqn:28}
\end{equation}
 gives the coupling of the virtual gluon  to the massive dipole (related to  $\sigma_{g^*d}$ in  Fig.1 by Mellin transform \cite{ca91,mu98}). 

In formula (\ref {eqn:ftl}),  $\Phi_{T}\equiv \vert\Psi^+_T \vert^2+\vert\Psi^-_T \vert^2$ (resp. $\Phi_{L}\equiv \vert\Psi_L \vert^2$)
are thus the probability distributions of $q \bar q$ configurations for transverse (resp. longitudinally) polarized photons. Note that there exists also subdominant BFKL contributions which depend \cite {ma99} on the interference term $2\Re e\left\{\Psi^+_T \left(\bar\Psi^-_T\right)\right\}.$ They play a r\^ole in the study of azimuthal correlations~\cite {del1,del2} beyond the leading BFKL terms. Introducing the Mellin-transforms of the photon probability distributions
 defined by:
\begin{equation}
\int \frac{d^2r}{2\pi} (rQ)^{2\!-\!2\gamma}\!\int dz
\>\Phi_{T,L}(r,z;Q^2)=\phi_{T,L}(\gamma)
\label{eqn:defphi}
\end{equation}
and performing
the integrations with respect to $r,z,$ the equation (\ref{eqn:ftl}) can be rewritten as
\beqa
\sigma^{\g}_{T,L}=
  \frac{32\pi^2}{Q^2}\int\frac{d\gamma}{2i\pi}
  \left(\frac Q{Q_t}\right)^{2\gamma}\!\! 
  e^{\frac{\alpha_sN_c}{\pi}\chi(\gamma)Y}\nonumber\\
\times \ \phi_{T,L}(\gamma)\ A_{el}(\gamma)
 \ .
\label{crosssection}
\eqa
This equation is identical to the BFKL expression of the same cross-sections   thanks to  the following identity \cite{mu98}
\begin{equation}
A_{el}(\gamma)\cdot\phi_{T,L}(\gamma)=\alpha_{em}\frac{\alpha_{s}N_c}{4\pi}
v(\gamma)\cdot \frac{V_{T,L}(\gamma)}{\gamma}\ ,
\label{eqn:relation}
\end{equation}
where
\beqa
V_T(\gamma)\!\!&=&\!\!\frac{\alpha_s}{6\pi} 
\frac{(1\!+\!\gamma)(2\!-\!\gamma)2^{1\!-\!2\gamma}}
{(1\!+\!2\gamma)(1\!-\!\frac{2}{3}\gamma)}
\frac{\Gamma(1\!+\!\gamma)\Gamma^3(1\!-\!\gamma)}{\Gamma(2\!-\!2\gamma)}
\nonumber \\ V_L(\gamma)\!\!&=&\!\!V_T(\gamma)\frac{2\gamma(1\!-\!\gamma)}{(1\!+\!\gamma)(2\!-\!\gamma)}
\label{eqn:coeft}
\eqa
are the $k$-factorized impact factors~\cite{ca91} for transverse and longitudinal photons.

One finally obtains the relation between the color singlet distribution functions and the related impact factors for a virtual photon as:
\begin{equation}
\phi_{T,L}(\gamma)\equiv\ \alpha_{em}e^2\ \frac{N_c}{4\pi\alpha_s} \frac {V_{T,L}(\gamma)}{\gamma }\ 
  \frac{1}{v(1-\gamma)}\ .
\label{eqn:phi}
\end{equation}
\noindent

The formula (\ref{eqn:phi}) gives the explicit relation between    the  high energy $k$ factorization
and the wave-function formalisms. 
The (resummed) perturbative QCD cross section can thus be
factorized equivalently in two ways: i) by the convolution of the photon gluon
cross section times the gluon coupling to the dipole (right hand side), 
ii) by the probability
distribution of a pair of quarks in the photon times the dipole-dipole elementary
interaction (left hand side). Note however, that ``soft'' targets  may lead to models differing only through the non perturbative extensions of the two formalisms, see, e.g.~\cite{mu98}.

\section{The gluon jet probe}

In order to determine the color singlet $q \bar q$ content of a gluon jet, we shall follow an approach very similar to the previous one for the photon probe.
Let us consider now the gluon-target cross-sections instead of the photon-target ones by writing
\beqa
\sigma^g_{T,L}=\int d^2r dz\;\int d^2r_t dz_t\ \times
\nonumber\\
\; 
\Phi^g_{T,L}(r,z;Q^2)\Phi^t(r_t,z_t;Q_t^2)\>\sigma_{d}(r,r_t;Y)\ ,
\label{eqn:gtl}
\eqa
and using the same equations (\ref{dipoles}-\ref{eqn:28}), one writes
\beqa
\sigma^g_{T,L}&=&
  \frac{32\pi^2}{Q^2}\int\frac{d\gamma}{2i\pi}
  \left(\frac Q{Q_t}\right)^{2\gamma}\!\! 
  e^{\frac{\alpha_sN_c}{\pi}\chi(\gamma)Y}\nn
&\times&\phi^g_{T,L}(\gamma)\ A_{el}(\gamma)
 \ ,
\label{crossssection}
\eqa
where now
\begin{equation}
\int \frac{d^2r}{2\pi} (Q r)^{2(1\!-\!\gamma)}\int dz
\>\Phi^g_{T,L}=\phi^g_{T,L}(\gamma)
\ ,
\label{eqn:defphi1}
\end{equation}
by definition.
The BFKL equation for the same cross-sections 
is particularly simple since the impact factors of the gluon are just {\it equal to one} by definition and the gluon coupling to the dipole is given by formula (\ref{eqn:28}). One thus get  the following identities 
\begin{equation}
A_{el}(\gamma)\cdot\phi^g_{T,L}(\gamma)\equiv\frac{\alpha_{s}^2}{4\pi} \frac {v(\gamma)}{(1\!-\!\gamma)}
\ ,
\label{eqn:relationg}
\end{equation}
where the factor $(1\!-\!\gamma)^{-1}$ comes from the integration of transverse momentum scales with lower bound $Q$ at the upper vertex (the unintegrated spectrum, appearing in the BFKL formalism, is expressed without this factor). 

Finally, using relation (\ref{eqn:27}) the result reads:
\begin{equation}
\phi^g_{T,L}(\gamma)\equiv \frac{1}{4\pi} \frac 1 {(1\!-\!\gamma)\ v(1\!-\!\gamma)}\ ,
\label{relationg}
\end{equation}
where the normalization of $\Phi^g_{T,L}(r,z;Q^2)$ is  $1$ after integration. This formula  is the basis of our derivation~\cite {pe99} of the gluon jet probe in terms of colorless $q\bar q$ configurations. All in all it consists in the fact that the reformulation of the impact factors of the gluon in terms of dipole configurations generate a definite prediction for the probability distribution of dipoles in a gluon probe.

It can be shown~\cite{pe99} that a  parametrization of the probability distributions of dipoles inside the gluon probe can be written as follows:
\begin{eqnarray}
\Phi^g_T \lr{z,r,Q^2}  &=& \left(z^2+(1\!-\!z)^2\right) \hat {Q}^2  \phi^g_T\lr{u} \\
\Phi^g_L \lr{z,r,Q^2}  &=&  4 z(1\!-\!z)\ \hat {Q}^2  \phi^g_L\lr{u} ,
\label{param}
\end{eqnarray}
where the functions $\phi^g_T$ and $\phi^g_L$ describe the configuration space features of the color singlet dipoles in the gluon probe and $u\equiv \hat Q r.$
These functions can be derived and lead to an interpretation in terms   of 
quantum states and wave-functions of these dipoles. Indeed, as discussed in~\cite{pe99}, the parametrization (\ref{param}) reflects the fact that the first-created dipole in the ordered in rapidity cascade of the QCD dipole model~\cite{mu94} brings almost the whole energy  and shares approximately the same energy between its quark  $z$ and antiquark $1\!-\!z,$ than the quark and antiquark directly coupled to the incident gluon, while the color had flown away . From this assumption (\ref{param}) and from the equation (\ref{relationg}), and after some algebra~\cite{pe99}, it is not too difficult to find the following solutions (valid within some approximation~\cite{pe99})
\begin{eqnarray}
&&\Phi_{T}^{g} \lr{z,r,Q^2}  \equiv \frac 2{\pi}\ \left( z^2\!+\!\lr{1\!- \!z}^2\right)\ \hat Q^2  \nonumber\\ 
&&\times\ \left\{\frac {J_1^2 \lr{u}}{\left(u\right)^2}+J_0^2 \lr{u}\!-\!J_1^2 \lr{u}\!-\!\frac {J_2^2 \lr{u}}{2\lr{u}^2}\right\}
 \nonumber\\
&&\Phi_{L}^{g} \lr{z,r,Q^2}   \equiv  \frac 6 {\pi} z\lr{1\!-\! z}\ \hat Q^2 
 \nonumber\\
 &&\times\ \left\{\frac {J_1^2 \lr{u}}{\lr{u}^2}\!+\!J_0^2 \lr{u}\!-\!\frac 23 J_1^2 \lr{u}\!-\!\frac 13 J_2^2 \lr{u}\right\}
,
\label{gluon}
\end{eqnarray}
 where $J_{0,1,2}$ are the usual Bessel functions of first kind. 
\section{gluon {\it versus} photon probes}

The resulting formulae (\ref{gluon}) call for some comments when compared to the photon wave functions (\ref{photon})   and  may have interesting 
implications for hard processes with forward jets.

{\bf i)} The dependence on the  parameter $u$ is much softer (power-like) for the gluon than for the photon (exponential). Thus, the size fluctuations of the dipoles in configuration space for a given scale $\hat {Q}$
are expected to be more important.

{\bf ii)} In the expression of the distributions $\Phi_{T,L}^{g},$ there are
a  terms with both signs.  Thus, contrary to the photon case, there cannot 
be interpreted as probability distributions . Indeed,
a formulation in terms of real and virtual intermediate dipole states
can match with formulae (\ref{gluon}). It amounts to consider that there are initial colorful states in the description of the gluon probe, since the gluon is not locally connected with a colorless state as is the photon. In practice \cite{pe99} this affects only the tail of the distributions in the
variable $u$ and one satisfies the psitivity condition on  the observable cross-sections after integration.

Both points {\bf i,ii)} may have interesting applications to the phenomenology of hard forward jets. On the one hand, it is expected to give 
specific predictions to the leading BFKL prediction for ``hard scattering''
cross-sections, since these predictions (especially for the small mass component
\cite{mu97}) depend quite strongly on the projectile wave functions in the 
configuration space. On the other hand, the next-leading and non perturbative corrections to the BFKL prediction could be different for the gluon w.r.t. the photon due to the larger size fluctuations obtained in the former case. This clearly requires further interesting studies.

{\bf Acknowledgements}
We thank St\'ephane Munier and Henri Navelet  for useful remarks and St\'efan Narison for the nice meeting in Montpellier.

\bigskip

{\bf Lev Lipatov} (St.Petersburg Nuclear Physics Institute, Russia)\ :

Do you consider the virtual gluon cross-section in the framework of the $k_T$-factorization, where the initial particle really is the reggeized gluon?

{\bf R. P.}\ :

 No, I consider a real gluon jet with large $k_T$ as for instance in the case of Mueller-Navelet jets~\cite {mu86}, and thus the ordinary renormalization-group factorization at short distance.

\bigskip

{\bf

\end{document}